\documentclass{appolb}
\usepackage{graphicx}

\begin{document}
\title{Hydrodynamic models of ultrarelativistic collisions 
\thanks{Presented by PB at XX Cracow Epiphany Conference on the Physics at the LHC,
8-10 January 2014}%
}
\author{Piotr Bo\.zek
\address{AGH University of Science and Technology, Faculty of Physics and Applied Computer Science, al. Mickiewicza 30, 30-059 Cracow, Poland\\
and  Institute of Nuclear Physics PAN, 31-342 Cracow, Poland}
\and
Wojciech Broniowski
\address{Institute of Nuclear Physics PAN, 31-342 Cracow, Poland\\
and Institute of Physics, Jan Kochanowski University, 25-406~Kielce, Poland}
}
\maketitle

\begin{abstract}
We discuss the use of the  hydrodynamic model for the description of 
the evolution of dense  matter formed in ultrarelativistic heavy-ion
 collisions. The collective flow observed in heavy-ion collisions at 
the BNL Relativistic Heavy Ion Collider and at the CERN Large Hadron Collider 
is consistent with the assumption that a fireball of strongly interacting 
matter is formed. Experimental results from p-Pb
 and d-Au collisions show  similar phenomena, which suggests
 that collective expansion appears in small systems as well. We 
review the recent application of the hydrodynamic model to small systems and 
discuss  limitations  and  possible further checks of  this scenario.
\end{abstract}
\PACS{25.75.-q,25.75.Dw,25.75.Nq}
  
\section{Introduction}

Nuclear collisions at ultrarelativistic energies 
lead to the formation of a dense fireball of quark-gluon plasma
\cite{Florkowski:2010zz,Heinz:2013th,Gale:2013da,Luzum:2013yya}.
The observation of the elliptic flow, the  asymmetry in the particle spectra 
between the in and out of the reaction plane directions,
is a strong evidence in favor  of the existence of  collective flow 
\cite{Back:2004je,Adams:2005dq,Adcox:2004mh, Aamodt:2010cz,Aad:2010bu,Chatrchyan:2011pb}. 
The interpretation of these experimental results assumes the expansion of 
a droplet of strongly interacting medium. 

If the systems evolves close to local thermodynamic equilibrium, 
relativistic hydrodynamic equations
\begin{equation}
\partial_\mu T^{\mu \nu}=0
\end{equation}
 can be used to describe the dynamics of the local energy density $\epsilon$, pressure $P$ and flow velocity $u^\mu$, where
the energy momentum tensor 
\begin{equation}
T^{\mu\nu} = (\epsilon+P) u^\mu u^\nu -P g^{\mu \nu} +\pi^{\mu\nu} + \Pi (g^{\mu \nu}- u^\mu u^\nu) \ .
\end{equation}
The stress tensor $\pi^{\mu \nu} $ and the bulk viscosity correction $\Pi$
are solutions of dynamical equation in the second order viscous 
hydrodynamic framework  \cite{IS,Song:2007ux,Chaudhuri:2006jd,Dusling:2007gi,Romatschke:2009im,Luzum:2008cw,Bozek:2009dw,Schenke:2010rr,Niemi:2011ix,Akamatsu:2013wyk,DelZanna:2013eua,Karpenko:2013wva}. 
The shear and bulk viscosity coefficients are important 
characteristics of the quark-gluon plasma. The estimated small value 
of the shear viscosity to entropy ratio $\eta/s$ is  not far from 
the AdS/CFT estimate $\eta/s=0.08$ 
\cite{Kovtun:2004de}, which
shows that the medium formed in heavy-ion collisions is strongly interacting.
 The extraction of the  shear viscosity coefficient is a difficult task,
as it requires the comparison of model calculations to experimental data on 
the elliptic and triangular flow at different collision centralities 
\cite{Romatschke:2007mq,Song:2008hj,Gale:2012rq,Luzum:2012wu}.
The problem comes from the uncertainty on the initial values of the spatial
ellipticity and triangularity. Another issue is related to the 
temperature dependence of $\eta/s$; in particular, $\eta/s$
 can be very different in the 
plasma and the hadronic phase \cite{Bozek:2009dw,Niemi:2011ix}.

The shape of the overlap region in the collision fluctuates from event to event and 
the eccentricity increases due to these fluctuations \cite{Alver:2007rm}. 
For each initial state, the hydrodynamic evolution is performed independently 
\cite{Schenke:2010rr,Petersen:2010cw,Gardim:2011xv,Bozek:2012fw,Qiu:2011hf,Niemi:2012aj,Pang:2012he}. The appearance of the triangular deformation from 
fluctuations brings in a qualitatively new observation, a non-zero triangular 
flow \cite{Alver:2010gr}. 
The initial density of the fireball is generated from a Monte Carlo model,
the Glauber model, f-KLN, IP-Glasma, URQMD,  or AMTP. The density and 
flow velocity evolves according to hydrodynamic equations and 
is driven by pressure gradients in the fireball. The collective 
expansion ends at the freeze-out hypersurface. 
That surface is usually defined as a constant temperature surface, or
equivalently as a cut-off in  local energy density. For smaller densities
 the collective expansion does not occur; individual hadrons are emitted 
from the freeze-out hypersurface. After freeze-out only hadron rescattering,
 resonance decay or creation can occur.
In the model calculations presented below we use 
Glauber Monte Carlo initial conditions \cite{Rybczynski:2013yba},
event by event hydrodynamic evolution with bulk and shear viscosity
 \cite{Bozek:2011ua}, and the THERMINATOR code to simulate statistical hadron emission at freeze-out and 
subsequent resonance decays
\cite{Kisiel:2005hn,Chojnacki:2011hb}.

\begin{figure}
\begin{center}
\includegraphics[angle=0,width=0.9 \textwidth]{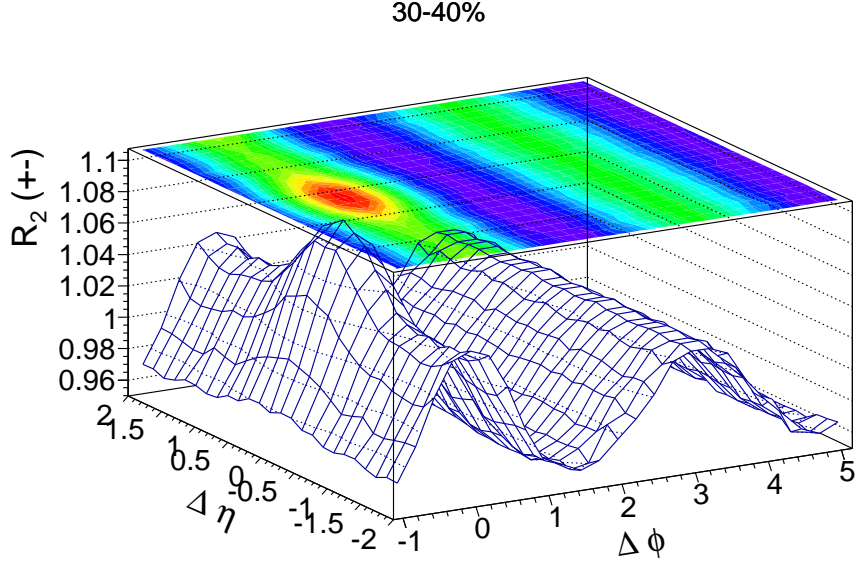}
\end{center}
\caption{Two particle correlations function ($R(\Delta \phi, \Delta \eta)=
C(\Delta \phi, \Delta \eta)$)
 in relative pseudorapidity and relative
azimuthal angle for unlike charged hadrons emitted  with $p_\perp>0.8$~GeV in $30-40$\% centrality 
 Au-Au collisions 
at $200$~GeV.  The calculation is based on event by event viscous hydrodynamics
with local charge correlations included. 
\label{fig:corr}} 
\end{figure}   

The fireball is elongated in the longitudinal direction (space-time rapidity) 
 and in the transverse direction it is deformed. The azimuthal deformation
 can be  effectively parametrized using $n-$order eccentricity
 coefficients
\begin{equation}
\epsilon_n e^{i n \phi_n}= \frac{\int \rho(x,y) e^{i n \phi  } r^n dx dy }
{\int \rho(x,y)  r^n dx dy } \ \ ,
\end{equation}
where $\phi_n$ defines the $n-$order event plane direction.
The transverse momentum spectra of emitted particles can be written as
\begin{eqnarray}
\frac{dN}{d^2_\perp dy}& =&\frac{dN}{2\pi p_\perp dp_\perp d\phi dy} \left( 
1+ 2 v_1 \cos(\phi - \psi_1) + 2 v_2 \cos(2(\phi - \psi_2)) \right. \nonumber \\
& &\left. + 2 v_3 \cos(3(\phi - \psi_3))+\dots \right) \ .
\end{eqnarray}
For $n=2$ and $3$ the hydrodynamic response is approximately linear 
\cite{Gardim:2011xv}
\begin{equation}
v_n \simeq A  \epsilon_n \ .
\end{equation}
The  hydrodynamic response $A$ depends on the details of
 the hydrodynamic evolution,  in particular,  it is sensitive to the  value 
of the shear viscosity.
The flow component in the  two-particle correlation 
function in relative azimuthal angle 
$\Delta \phi$ and relative 
pseudorapidity $\Delta \eta$  is approximately independent of $\Delta \eta$, while its  harmonic expansion  in azimuthal angle is
 given by the flow coefficients
\begin{equation}
C(\Delta \phi , \Delta \eta) \propto  1 + 2 v_1^2 \cos(\Delta \phi) + 2 v_2^2 \cos(2 \Delta \phi) + 2 v_3^2 \cos(3\Delta  \phi) + \dots \ .
\label{eq:corr}
\end{equation}
The two-dimensional plot of the correlation function has a same-
($\Delta \phi \simeq 0$) and away-side side ($\Delta \phi \simeq \pi$) ridge
(Fig.~\ref{fig:corr}).
Non-flow correlations contribute to the two-particle correlation $C(\Delta \phi , \Delta \eta)$: the jet-like correlations, resonance decays, and the local 
charge conservation  at small $\Delta\eta$ and $\Delta \phi$~\cite{Bozek:2012en},
and transverse momentum conservation in the away-side ridge region \cite{Borghini:2000cm}.
Experimental estimates of the flow coefficients from the second order cumulants 
are equivalent to an event average of the two-particle correlation function 
$ \langle v_n^2 \rangle = \langle C(\Delta \phi) \cos(n \Delta \phi) \rangle $,
which sums up the flow fluctuations as well as the the average flow.

The space-time pattern of particle emission from the expanding fluid can be 
extracted from same particle interferometry correlations 
\cite{Wiedemann:1999qn,Lisa:2005dd}. The interferometry radii measure the size
of the emission region for  pairs of particles of a given momentum \cite{Akkelin:1995gh}.  The value of the femtoscopy 
radii serves as an estimate of the size of the fireball at freeze-out,
and it is consistent with the size of the initial fireball
assumed in hydrodynamic models, supplemented with the increase during  
the collective expansion phase. The reduction of the interferometry
radii with  the average momentum of the pair  indicates a strong correlation 
between the flow and position. This correlation can be reproduced
in hydrodynamic calculations when a realistic, hard equation of state 
is used~\cite{Broniowski:2008vp,Pratt:2008qv,Bozek:2010er,Karpenko:2012yf}.

\section{Collective flow in small systems}

Experimental results from relativistic heavy-ion  collisions present
strong evidence for the formation of a dense fireball that expands collectively.
Ultrarelativistic d-Au and p-Pb collisions have been performed in order to 
study phenomena unrelated to plasma formation and to obtain reference data
for heavy-ion experiments \cite{Salgado:2011pf}.
On the other hand, extrapolations of the initial 
 energy density from peripheral Pb-Pb to p-Pb collisions indicate that 
collective expansion could take place in p-Pb collisions at the LHC.
The hydrodynamic model predicts a significant transverse expansion of
 the fireball formed in high multiplicity  p-Pb collisions \cite{Bozek:2011if}.

\begin{figure}
\begin{center}
\includegraphics[angle=0,width=0.900 \textwidth]{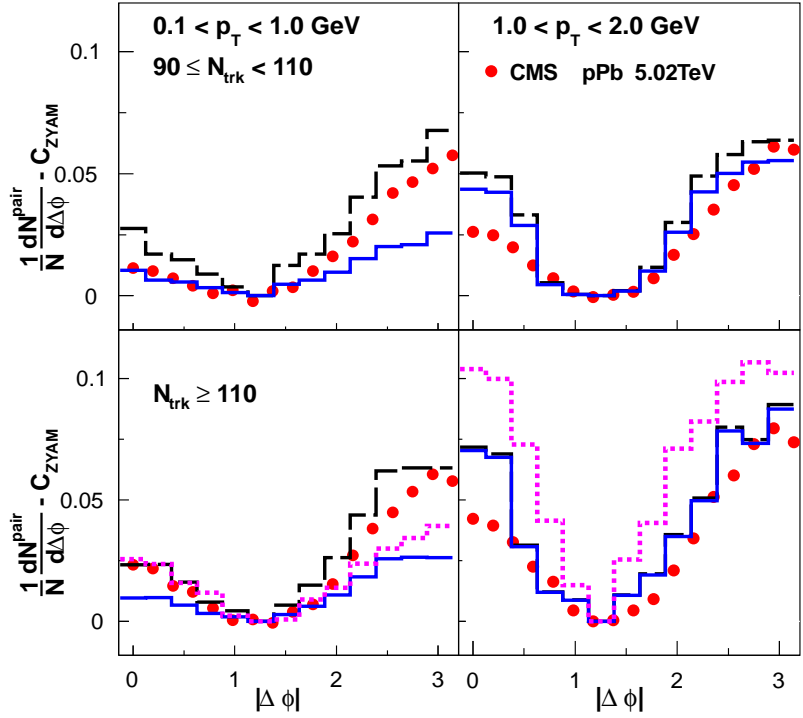}
\end{center}
\caption{The   ZYAM  subtracted correlation function   in  p-Pb collisions. 
The CMS measurement~\cite{CMS:2012qk} is shown as filled circles.  
The results 
of the hydrodynamic model with the normalized correlation functions 
are shown with the solid lines. 
The dashed lines represent the ZYAM subtracted results of the hydrodynamic
 model and no rescaling. The dotted lines show the results obtained with on initial time of $0.2$  instead of $0.6$~fm/c (from \cite{Bozek:2012gr}).
\label{fig:corrpanel}} 
\end{figure}

The observed 
two-particle correlation functions in p-Pb collisions 
 \cite{CMS:2012qk,Abelev:2012ola,Aad:2012gla} are qualitatively 
similar to the A-A case, as
two ridge like structures elongated in the pseudorapidity direction are 
clearly visible. These structures can be explained as due to the collective flow
and the transverse momentum conservation \cite{Bozek:2012gr}. The
elliptic and triangular  collective flow components, together with the 
$\cos(\Delta\phi)$ contribution from momentum conservation, qualitatively
reproduce the observed  projected correlation function $C(\Delta \phi)$ 
(Fig.~\ref{fig:corrpanel}). In the one-dimensional correlation 
function presented in  Fig.~\ref{fig:corrpanel} the short range non-flow correlations are reduced using a cut
 $|\Delta \eta|>2$ in the projection.  We note that a similar mechanism could explain 
the ridge structures observed in the high multiplicity p-p collisions \cite{Khachatryan:2010gv}, 
but  definite conclusions are more difficult here due to 
stronger non-flow contributions \cite{Bozek:2010pb}. 
The observed two ridge structure of the correlation function can arise 
due to initial state effects
\cite{Dusling:2012iga,Dusling:2012cg,Dusling:2013oia,Dusling:2012wy},
leading an enhancement of the gluon emission at small angles. It is important to be able to disentangle the two scenarios.

\begin{figure}[h]
\begin{center}
\includegraphics[width=0.800 \textwidth]{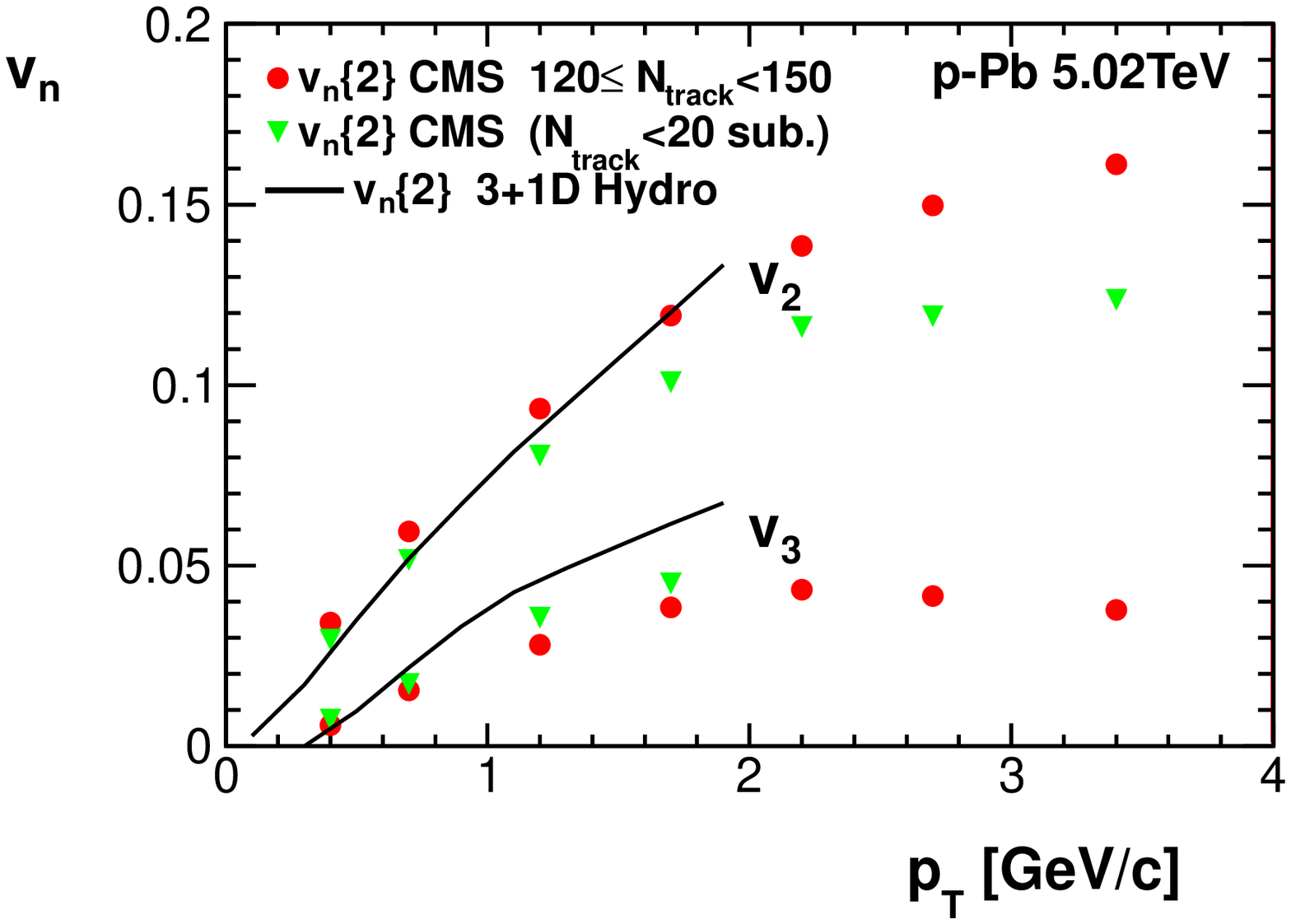}
\end{center}
\caption{\label{fig:v23}   $v_2$ and $v_3$ for charged 
particles from the hydrodynamic calculation compared to 
 CMS Collaboration data  ~\cite{Chatrchyan:2013nka} (from \cite{Bozek:2013ska}).}
\end{figure}

The extraction of the flow coefficients $v_2$ and $v_3$ in 
p-Pb collisions is difficult  due to significant non-flow correlations.
Methods involving subtraction of peripheral from central correlation functions, employing the rapidity gap, or 
higher order cumulants can be used for that purpose
 \cite{Aad:2013fja,Chatrchyan:2013nka,ABELEV:2013wsa}.
The elliptic and triangular flow of charged particles  
in high multiplicity p-Pb events is well described 
by the hydrodynamic model \cite{Bozek:2013uha} (Fig. \ref{fig:v23}).
Qualitatively similar results are obtained in hydrodynamic calculations using
various assumptions about the initial density 
\cite{Qin:2013bha,Bzdak:2013zma,Werner:2013ipa,Nagle:2013lja}.
In p-Pb collisions the initial density is formed from
  a small number of independent sources. This leads to the approximate
 equality of eccentricities from higher order cumulants
$v_2\{ 4\}\simeq v_2\{ 6\} \simeq v_2\{ 8\}$ 
\cite{Bzdak:2013rya,Yan:2013laa,Bzdak:2013raa}.

 The fireball 
is smaller and lives shorter than in A-A collisions. It makes 
the quantitative prediction of the hydrodynamic model more sensible to 
the assumed initial state scenario or to 
changes in phenomenological  parameters.
The shape of the fireball depends on the modeling of the energy deposition 
on small scales and should be described using subnuclear degrees of freedom 
\cite{Bozek:2013uha,Bzdak:2013zma}. The amount of the transverse 
flow generated changes noticeably when  the initial thermalization 
time or  the 
freeze-out density are lowered.
 More importantly, we should be aware that the
 applicability of second order viscous hydrodynamics is less justified 
in small systems, when
velocity gradients are large.

\begin{figure}[h]
\begin{center}
\includegraphics[width=0.800 \textwidth]{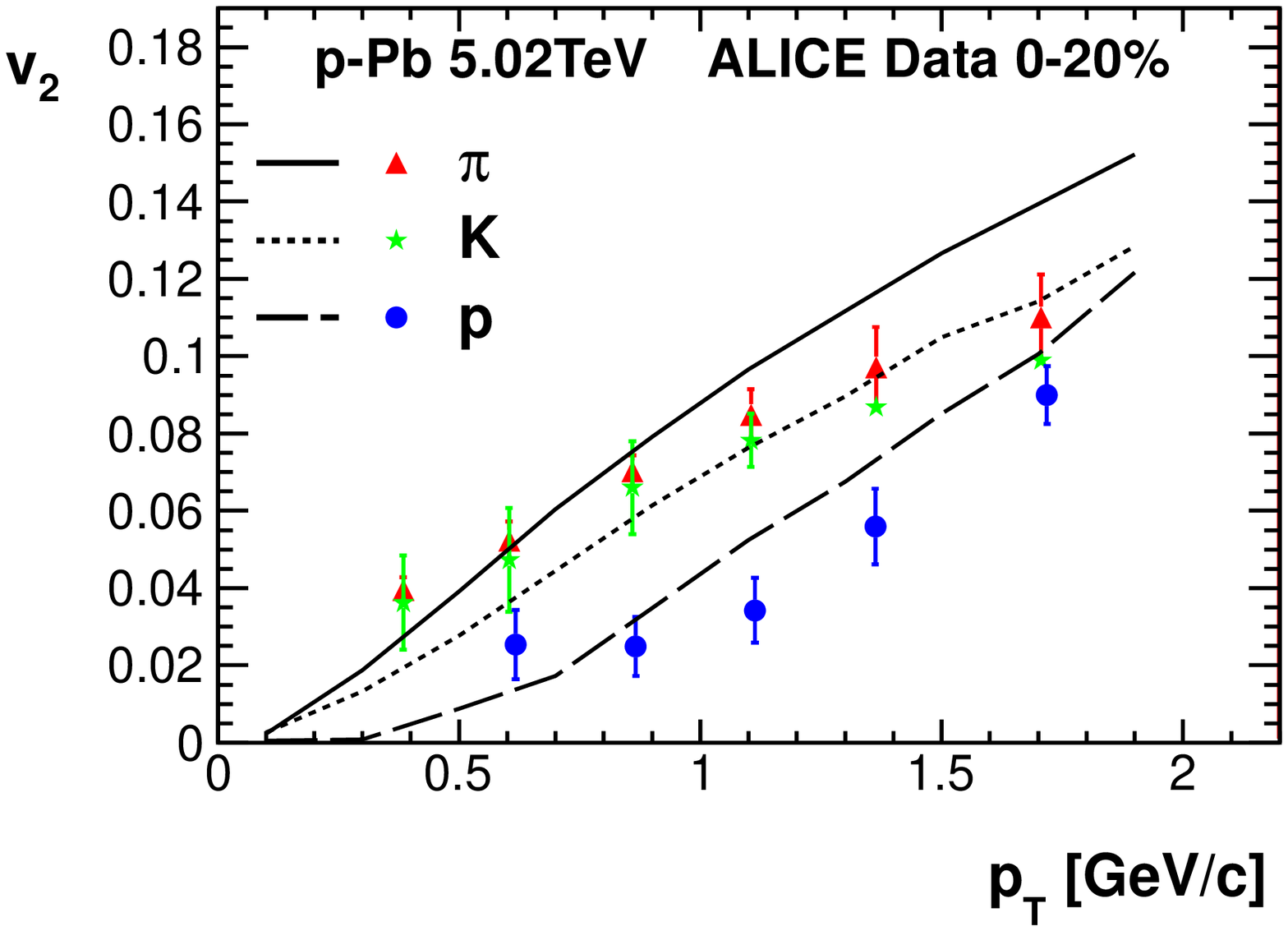}
\end{center}
\caption{\label{fig:v2id}  $v_2(p_\perp)$ for pions, kaons, 
and protons from the hydrodynamic model, compared to
ALICE Collaboration data~\cite{ABELEV:2013wsa} (from \cite{Bozek:2013ska}).}
\end{figure}

The elliptic flow coefficient as function of transverse momentum, $v_2(p_\perp)$,  
splits for different particles. In particular, the elliptic 
flow of pions is larger than for protons, for $p_\perp < 1.5$~GeV. This 
appears in hydrodynamic models as the  mass splitting of the elliptic flow.
The results for the elliptic flow of identified particles reproduce 
qualitatively  the experimental pion-proton splitting 
\cite{Bozek:2013ska} (Fig. \ref{fig:v2id}).

\begin{figure}[h]
\begin{center}
\includegraphics[width=0.800 \textwidth]{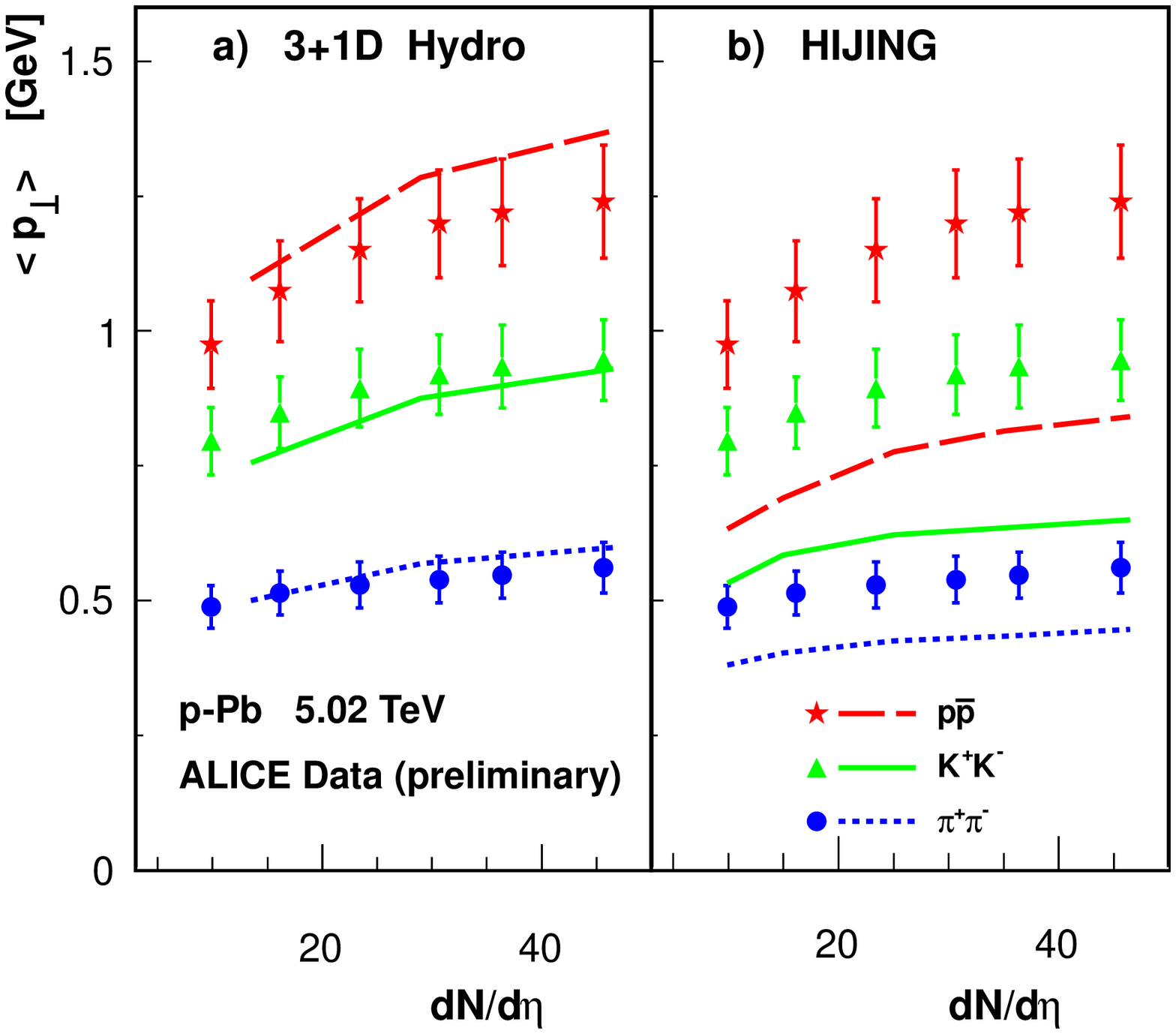}
\end{center}
\caption{\label{fig:avpt}  Average transverse momentum for pions, kaons, 
and protons in p-Pb collisions from the hydrodynamic model (left panel) and from the HIJING model (right panel),
 compared to
ALICE Collaboration data~\cite{Abelev:2013haa} (from \cite{Bozek:2013ska}).}
\end{figure}

The momentum of 
particles emitted from a moving fluid element  
gets a contribution from the collective velocity, it is bigger for 
massive particles. This yields a mass hierarchy in the average 
transverse momentum  of particles \cite{Kolb:2000sd}. Scenarios without 
collective expansion can predict the increase of the average transverse flow 
with particle mass in p-p collisions 
\cite{Ortiz:2013yxa},  in accordance with experimental results,
but models based on the convolution of independent nucleon-nucleon collisions
cannot reproduce the experimental results in p-Pb interactions 
\cite{Bzdak:2013lva}. An example of the average transverse momentum from a superposition model (HIJING) is given in the right panel of Fig. \ref{fig:avpt},
where the value of $\langle p_\perp \rangle$ and its mass splitting are smaller than 
in the experiment. A hydrodynamic calculation \cite{Bozek:2013ska} 
can reproduce the mass hierarchy of the average transverse momenta 
(Fig. \ref{fig:avpt}, left panel). The rapidity dependence of the 
average transverse momentum could serve as a way to disentangle between 
the collective expansion and color glass condensate scenarios 
\cite{Bozek:2013sda}. In the hydrodynamic model the transverse push is smaller when going to the proton side, while the reverse is true in the color glass 
condensate approach.

The scenario based on  the formation and expansion of a dense fireball
brings the question about the possibility to measure its size. The
 interferometry radii could measure the size of the fireball at freeze-out and
 its momentum dependence could provide  information on the transverse flow.
If the  initial size of the fireball increases during the expansion, the femtoscopy radii measured in p-Pb collisions should be larger than in p-p interaction
\cite{Bozek:2013df}. The initial size of the interaction region is small in the IP-glasma scenario \cite{Bzdak:2013zma}, 
hence measuring such small radii in the 
experiment would indicate that the expansion does not happen.

The collective flow in d-A collisions is another  interesting possibility. It has 
been noted that the ellipticity in high multiplicity d-A collisions is big, 
with the central (highest-multiplicity) events corresponding to the deuteron hitting
the bigger nucleus  side-wise 
\cite{Bozek:2011if}.
An intrinsic  deformation of the  fireball appears, unlike for p-Pb collisions 
where the deformation is entirely due to fluctuations. The two-particle 
correlation functions in the d-Au collisions at $200$~GeV have been analyzed 
using similar methods as for the p-Pb collisions at the LHC energies
 \cite{Adare:2013piz}. The extracted elliptic flow coefficient is large  
as expected from hydrodynamic calculations, while the triangular flow is negligible.

\section{Discussion}

\subsection{Summary of results}

The relativistic heavy-ion research program has provided a strong 
evidence for the formation of strongly interacting quark gluon plasma in 
A-A collisions.
Recent experimental results indicate that final state interaction followed with collective evolution could 
also be important in ultrarelativistic collisions of small on large systems, p-Pb at $2.76$~TeV
and d-Au collisions at $200$~GeV. 
Observations favoring this scenario are:
\begin{itemize}
\item The observation of elliptic and triangular flow in p-Pb collisions
\cite{Aad:2013fja,Chatrchyan:2013nka,ABELEV:2013wsa},
consistent with model calculations 
\cite{Bozek:2013uha,Qin:2013bha,Bzdak:2013zma,Werner:2013ipa,Nagle:2013lja},
\item  An even larger elliptic flow in d-Au collisions  \cite{Adare:2013piz},
 in line with 
hydrodynamic predictions \cite{Bozek:2011if},
\item The mass hierarchy of the elliptic flow 
and of the  average transverse flow 
\cite{ABELEV:2013wsa,Abelev:2013haa,Chatrchyan:2013eya}
\item Similarity between p-Pb collisions and peripheral Pb-Pb collisions
\cite{Chatrchyan:2013nka,Basar:2013hea}.
\end{itemize}

The experimental results listed in the previous section motivate further 
studies:
\begin{itemize}
\item The analysis of the proton-proton and peripheral A-A collision carried out in a similar way as in 
p-A and d-A collisions. Such a program could be used to find the possible onset of collectivity in small systems.  
The study of the  onset  of
 jet quenching 
in very peripheral A-A and p-A collisions would give additional information
 on the nature of the matter formed in small systems.
 \item The energy scan of  p-A and d-A collisions to find the onset of collectivity as a function of the energy density.
\item Experimental studies of collisions of small deformed projectiles on big nuclei. This  subject is discussed  in the next subsection.
\end{itemize}

\subsection{Why small on big collisions?}

\begin{figure}[h]
\begin{center}
\includegraphics[width=0.800 \textwidth]{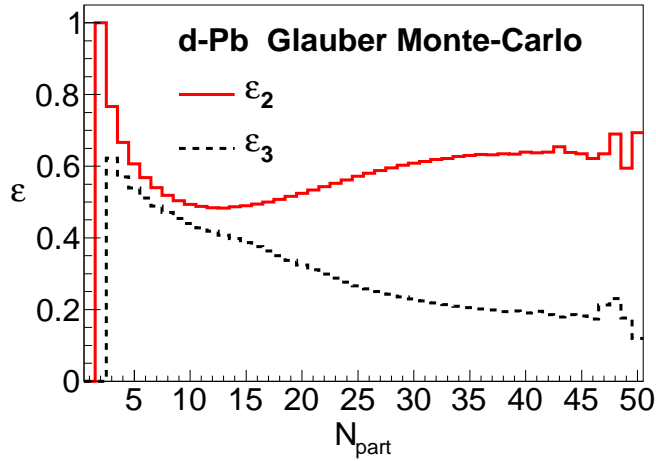}
\end{center}
\caption{\label{fig:dA}  The ellipticity and triangularity in d-Pb collisions 
as function of the number of participants (from \cite{Bozek:2011if}).}
\end{figure}

We now briefly discuss motivations for performing experiments 
with a very asymmetric projectile and target, other than to 
study various aspects of the initial state dynamics
\cite{Salgado:2011pf}. In view of the results indicating that collective flow
appears in such collisions, it is important to study such systems in more 
detail.
The extraction of  flow coefficients is difficult because 
of significant non-flow contributions. It should be kept in mind that 
alternative scenarios based on the color glass condensate approach are used to interpret the observations.
Therefore further experiments are needed to validate or disprove the 
collectivity in small systems.

A mechanism to control the eccentricity has been
discussed for d-A collisions \cite{Bozek:2011if}. 
The collision of a  deformed projectile 
with a large nucleus can be viewed as a small deformed (in this case a dumbbell shaped) nucleus hitting a wall. 
The orientation of the projectile determines the ellipticity of the 
fireball. By triggering on high multiplicity events we choose collisions
where the projectile makes the largest damage when colliding, i.e., events with a large number of participants. The configurations 
 relevant in this case are those where the deuteron 
hits the larger nucleus side-wise.
Thus we expect the largest ellipticity for the most central collisions
(Fig. \ref{fig:dA}). The PHENIX collaboration indeed observes 
a large $v_2$ in d-Au collisions  \cite{Adare:2013piz}, 
consistent with such estimates. A larger nucleus with a quadrupole 
deformation, such as $^9$Be,  could be used instead of the deuteron.

When using a small deformed projectile of triangular shape  \cite{Sickles:2013mua}, 
such as triton or $^3$He, a significant triangular flow should appear. 
Hydrodynamic simulations show  that $v_3$ in $^3$He-Au is larger than
in d-Au or p-Au collisions \cite{Nagle:2013lja}. 

\begin{figure}
\begin{center}
\includegraphics[angle=0,width=0.6 \textwidth]{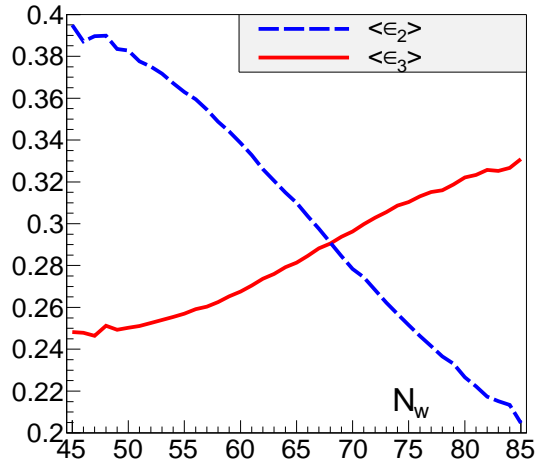}
\end{center}
\caption{Glauber-model prediction for the event-averaged ellipticity and triangularity of the initial fireball created in the ${}^{12}$C-${}^{208}$Pb
collisions at the highest RHIC energies of $\sqrt{s_{\rm NN}}=200$~GeV. We use the mixed Glauber model with $\sigma_{\rm NN}=42$~mb and the 
fraction of the binary collisions $a=0.145$.
\label{fig:c12}} 
\end{figure}

A very promising and interesting systems to study the triangularity of the projectile is to use  $^{12}$C \cite{Broniowski_pub:2013dia}. Nucleons in 
the ground state of the $^{12}$C nucleus are strongly clustered into three $\alpha$
particles (see,
e.g.,~\cite{blatt19521952theoretical} for an early review,~\cite{brink2008history} for some history, 
or~\cite{brink1965alpha,freer2007clustered,ikeda2010clusters,beck2012clusters} for references). For ${}^{12}$C-${}^{208}$Pb events with a 
large number of participants, the triangularity
can be as high as $0.3$ (Fig. \ref{fig:c12}). Moreover, 
due to the intrinsic triangular shape of the carbon, the triangularity of the fireball 
increases with the number of participants (in analogy to the ellipticity in the d-A case shown in Fig.~\ref{fig:dA}), providing a vivid qualitative signal of the 
clusterization. We note that
the size of the 
interaction region and the multiplicity are much larger than in $^3$He-Au 
collisions, ranging up to 85 wounded nucleons for the highest RHIC energy. 
Thus the collective scenario is 
anticipated for the $^{12}$C-Au 
collisions. Finally, we stress that from the quantum-mechanical point view, the flow analysis would
present a unique way to get snapshots of the intrinsic wave function of 
the carbon nucleus at the instant of the collision.

Performing a series of experiments using small projectiles
hitting a large nucleus would clarify the role of the final state interactions.
If the collective expansion is valid, one expects a moderate $v_2$ with 
a smaller $v_3$ in p-A collisions,  a large $v_2$ and negligible $v_3$ in d-A or $^9$Be-A collisions, and a large $v_3$ and  $v_2$ in $^3$He-A or $^{12}$C-A collisions, 
with specific correlations with multiplicity~\cite{Broniowski_pub:2013dia}.

\subsection{Limits of collectivity}

Experimental indications of collective expansion in small systems rise 
the question about the limits  of applicability of the hydrodynamic model.
At the very early stage of the collision the system evolves far from 
equilibrium. A strong pressure asymmetry is expected between the 
longitudinal and transverse direction. Since the total evolution 
time is shorter for p-A collisions, one hopes to be able  to investigate 
this early transient stage. Quantitative predictions require going beyond the 
framework of the near-equilibrium relativistic hydrodynamics
 \cite{Florkowski:2013sk,Heller:2011ju}. 

If the gradients of the transverse velocity get big in a small system, the
hydrodynamic approach would cease to be valid even for the transverse 
expansion. This means that the second order viscous hydrodynamics breaks down. 
Phenomenologically, as long as the scale at which hadrons are formed is 
smaller than the size of the system, local correlations between flow and 
position (collective flow) can appear. It remains a challenge to provide a theoretical framework able to give quantitative predictions in that case.

\section*{Acknowledgments}
This work is partly supported by the National Science Centre, 
Poland, grants DEC-2012/05/B/ST2/02528 and DEC-2012/06/A/ST2/00390, and PL-Grid infrastructure.

\bibliography{../../../hydr,../../../hydr_wb,../../../clusters}

\begin{thebibliography}{10}
\expandafter\ifx\csname url\endcsname\relax
  \def\url#1{\texttt{#1}}\fi
\expandafter\ifx\csname urlprefix\endcsname\relax\def\urlprefix{URL }\fi

\bibitem{Florkowski:2010zz}
W.~Florkowski, {\em {Phenomenology of Ultra-Relativistic Heavy-Ion
  Collisions}}, World Scientific Publishing Company, Singapore, 2010.

\bibitem{Heinz:2013th}
U.~Heinz, R.~Snellings, Ann.Rev.Nucl.Part.Sci., {\bf 63} (2013) 123, 1301.2826.

\bibitem{Gale:2013da}
C.~Gale, S.~Jeon, B.~Schenke, Int.J.Mod.Phys., {\bf A28} (2013) 1340011.

\bibitem{Luzum:2013yya}
M.~Luzum, H.~Petersen, 1312.5503.

\bibitem{Back:2004je}
B.~B. Back, {\it et~al.}, PHOBOS, Nucl. Phys., {\bf A757} (2005) 28.

\bibitem{Adams:2005dq}
J.~Adams, {\it et~al.}, STAR Collaboration, Nucl. Phys., {\bf A757} (2005) 102.

\bibitem{Adcox:2004mh}
K.~Adcox, {\it et~al.}, PHENIX Collaboration, Nucl. Phys., {\bf A757} (2005)
  184.

\bibitem{Aamodt:2010cz}
K.~Aamodt, {\it et~al.}, ALICE Collaboration, Phys. Rev. Lett., {\bf 106}
  (2011) 032301.

\bibitem{Aad:2010bu}
G.~Aad, {\it et~al.}, ATLAS, Phys. Rev. Lett., {\bf 105} (2010) 252303.

\bibitem{Chatrchyan:2011pb}
S.~Chatrchyan, {\it et~al.}, CMS, JHEP, {\bf 1108} (2011) 141.

\bibitem{IS}
W.~Israel, J.~Stewart, Annals Phys., {\bf 118} (1979) 341.

\bibitem{Song:2007ux}
H.~Song, U.~W. Heinz, Phys. Rev., {\bf C77} (2008) 064901.

\bibitem{Chaudhuri:2006jd}
A.~K. Chaudhuri, Phys. Rev., {\bf C74} (2006) 044904.

\bibitem{Dusling:2007gi}
K.~Dusling, D.~Teaney, Phys. Rev., {\bf C77} (2008) 034905.

\bibitem{Romatschke:2009im}
P.~Romatschke, Int. J. Mod. Phys., {\bf E19} (2010) 1.

\bibitem{Luzum:2008cw}
M.~Luzum, P.~Romatschke, Phys. Rev., {\bf C78} (2008) 034915.

\bibitem{Bozek:2009dw}
P.~Bo\.zek, Phys. Rev., {\bf C81} (2010) 034909.

\bibitem{Schenke:2010rr}
B.~Schenke, S.~Jeon, C.~Gale, Phys. Rev. Lett., {\bf 106} (2011) 042301.

\bibitem{Niemi:2011ix}
H.~Niemi, G.~S. Denicol, P.~Huovinen, E.~Molnar, D.~H. Rischke, Phys.Rev.Lett.,
  {\bf 106} (2011) 212302.

\bibitem{Akamatsu:2013wyk}
Y.~Akamatsu, S.-I. Inutsuka, C.~Nonaka, M.~Takamoto, J.Comput.Phys., {\bf 256}
  (2014) 34.

\bibitem{DelZanna:2013eua}
L.~Del~Zanna, V.~Chandra, G.~Inghirami, V.~Rolando, A.~Beraudo, {\it et~al.},
  Eur.Phys.J., {\bf C73} (2013) 2524.

\bibitem{Karpenko:2013wva}
I.~Karpenko, P.~Huovinen, M.~Bleicher, 1312.4160.

\bibitem{Kovtun:2004de}
P.~K. Kovtun, D.~T. Son, A.~O. Starinets, Phys. Rev. Lett., {\bf 94} (2005)
  111601.

\bibitem{Romatschke:2007mq}
P.~Romatschke, U.~Romatschke, Phys. Rev. Lett., {\bf 99} (2007) 172301.

\bibitem{Song:2008hj}
H.~Song, U.~W. Heinz, J. Phys., {\bf G36} (2009) 064033.

\bibitem{Gale:2012rq}
C.~Gale, S.~Jeon, B.~Schenke, P.~Tribedy, R.~Venugopalan, Phys.Rev.Lett., {\bf
  110} (2013) 012302.

\bibitem{Luzum:2012wu}
M.~Luzum, J.-Y. Ollitrault, Nucl.Phys.A904-905, {\bf 2013} (2013) 377c.

\bibitem{Alver:2007rm}
B.~Alver, {\it et~al.}, PHOBOS, J. Phys., {\bf G34} (2007) S907.

\bibitem{Petersen:2010cw}
H.~Petersen, G.-Y. Qin, S.~A. Bass, B.~Muller, Phys. Rev., {\bf C82} (2010)
  041901.

\bibitem{Gardim:2011xv}
F.~G. Gardim, F.~Grassi, M.~Luzum, J.-Y. Ollitrault, Phys. Rev., {\bf C85}
  (2012) 024908.

\bibitem{Bozek:2012fw}
P.~Bo\.zek, W.~Broniowski, Phys. Rev., {\bf C85} (2012) 044910.

\bibitem{Qiu:2011hf}
Z.~Qiu, C.~Shen, U.~Heinz, Phys. Lett., {\bf B707} (2012) 151.

\bibitem{Niemi:2012aj}
H.~Niemi, G.~Denicol, H.~Holopainen, P.~Huovinen, Phys. Rev., {\bf C87} (2013)
  054901.

\bibitem{Pang:2012he}
L.~Pang, Q.~Wang, X.-N. Wang, Phys. Rev., {\bf C86} (2012) 024911.

\bibitem{Alver:2010gr}
B.~Alver, G.~Roland, Phys. Rev., {\bf C81} (2010) 054905.

\bibitem{Rybczynski:2013yba}
M.~Rybczynski, G.~Stefanek, W.~Broniowski, P.~Bo\.zek, 1310.5475.

\bibitem{Bozek:2011ua}
P.~Bo\.zek, Phys. Rev., {\bf C85} (2012) 034901.

\bibitem{Kisiel:2005hn}
A.~Kisiel, T.~{Ta\l{}u\'c}, W.~Broniowski, W.~Florkowski, Comput. Phys.
  Commun., {\bf 174} (2006) 669.

\bibitem{Chojnacki:2011hb}
M.~Chojnacki, A.~Kisiel, W.~Florkowski, W.~Broniowski, Comput. Phys. Commun.,
  {\bf 183} (2012) 746.

\bibitem{Bozek:2012en}
P.~Bo\.zek, W.~Broniowski, Phys. Rev. Lett., {\bf 109} (2012) 062301.

\bibitem{Borghini:2000cm}
N.~Borghini, P.~M. Dinh, J.-Y. Ollitrault, Phys. Rev., {\bf C62} (2000) 034902.

\bibitem{Wiedemann:1999qn}
U.~A. Wiedemann, U.~W. Heinz, Phys. Rept., {\bf 319} (1999) 145.

\bibitem{Lisa:2005dd}
M.~A. Lisa, S.~Pratt, R.~Soltz, U.~Wiedemann, Ann. Rev. Nucl. Part. Sci., {\bf
  55} (2005) 357.

\bibitem{Akkelin:1995gh}
S.~Akkelin, Y.~Sinyukov, Phys. Lett., {\bf B356} (1995) 525.

\bibitem{Broniowski:2008vp}
W.~Broniowski, M.~Chojnacki, W.~Florkowski, A.~Kisiel, Phys. Rev. Lett., {\bf
  101} (2008) 022301.

\bibitem{Pratt:2008qv}
S.~Pratt, Phys. Rev. Lett., {\bf 102} (2009) 232301.

\bibitem{Bozek:2010er}
P.~Bo\.zek, Phys. Rev., {\bf C83} (2011) 044910.

\bibitem{Karpenko:2012yf}
I.~Karpenko, Y.~Sinyukov, K.~Werner, Phys. Rev., {\bf C87} (2013) 024914.

\bibitem{Salgado:2011pf}
C.~A. Salgado, J.Phys.G, {\bf G38} (2011) 124036.

\bibitem{Bozek:2011if}
P.~Bo\.zek, Phys. Rev., {\bf C85} (2012) 014911.

\bibitem{CMS:2012qk}
S.~Chatrchyan, {\it et~al.}, CMS Collaboration, Phys. Lett., {\bf B718} (2013)
  795.

\bibitem{Abelev:2012ola}
B.~Abelev, {\it et~al.}, ALICE Collaboration, Phys. Lett., {\bf B719} (2013)
  29.

\bibitem{Aad:2012gla}
G.~Aad, {\it et~al.}, ATLAS Collaboration, Phys. Rev. Lett., {\bf 110} (2013)
  182302.

\bibitem{Bozek:2012gr}
P.~Bo\.zek, W.~Broniowski, Phys. Lett., {\bf B718} (2013) 1557.

\bibitem{Khachatryan:2010gv}
V.~Khachatryan, {\it et~al.}, CMS, JHEP, {\bf 09} (2010) 091.

\bibitem{Bozek:2010pb}
P.~Bo\.zek, Eur.Phys.J., {\bf C71} (2011) 1530.

\bibitem{Dusling:2012iga}
K.~Dusling, R.~Venugopalan, Phys. Rev. Lett., {\bf 108} (2012) 262001.

\bibitem{Dusling:2012cg}
K.~Dusling, R.~Venugopalan, Phys. Rev. D, {\bf 87} (2013) 051502.

\bibitem{Dusling:2013oia}
K.~Dusling, R.~Venugopalan, Phys. Rev., {\bf D87} (2013) 094034.

\bibitem{Dusling:2012wy}
K.~Dusling, R.~Venugopalan, Phys. Rev. D, {\bf 87} (2013) 054014.

\bibitem{Aad:2013fja}
G.~Aad, {\it et~al.}, ATLAS Collaboration, Phys. Lett., {\bf B725} (2013) 60.

\bibitem{Chatrchyan:2013nka}
S.~Chatrchyan, {\it et~al.}, CMS Collaboration, Phys. Lett., {\bf B724} (2013)
  213.

\bibitem{ABELEV:2013wsa}
B.~B. Abelev, {\it et~al.}, ALICE Collaboration, Phys. Lett., {\bf B726} (2013)
  164.

\bibitem{Bozek:2013uha}
P.~Bo\.zek, W.~Broniowski, Phys. Rev., {\bf C88} (2013) 014903.

\bibitem{Qin:2013bha}
G.-Y. Qin, B.~M{\"u}ller, 1306.3439.

\bibitem{Bzdak:2013zma}
A.~Bzdak, B.~Schenke, P.~Tribedy, R.~Venugopalan, Phys. Rev., {\bf C87} (2013)
  064906.

\bibitem{Werner:2013ipa}
K.~Werner, M.~Bleicher, B.~Guiot, I.~Karpenko, T.~Pierog, 1307.4379.

\bibitem{Nagle:2013lja}
J.~Nagle, A.~Adare, S.~Beckman, T.~Koblesky, J.~O. Koop, {\it et~al.},
  1312.4565.

\bibitem{Bzdak:2013rya}
A.~Bzdak, P.~Bozek, L.~McLerran, 1311.7325.

\bibitem{Yan:2013laa}
L.~Yan, J.-Y. Ollitrault, Phys.Rev.Lett., {\bf 112} (2014) 082301.

\bibitem{Bzdak:2013raa}
A.~Bzdak, V.~Skokov, 1312.7349.

\bibitem{Bozek:2013ska}
P.~Bo\.zek, W.~Broniowski, G.~Torrieri, Phys. Rev. Lett., {\bf 111} (2013)
  172303.

\bibitem{Abelev:2013haa}
B.~B. Abelev, {\it et~al.}, ALICE Collaboration, 1307.6796.

\bibitem{Kolb:2000sd}
P.~F. Kolb, J.~Sollfrank, U.~W. Heinz, Phys. Rev., {\bf C62} (2000) 054909.

\bibitem{Ortiz:2013yxa}
A.~Ortiz, P.~Christiansen, E.~Cuautle, I.~Maldonado, G.~Paic, Phys. Rev. Lett.,
  {\bf 111} (2013) 042001.

\bibitem{Bzdak:2013lva}
A.~Bzdak, V.~Skokov, Phys. Lett., {\bf B726} (2013) 408.

\bibitem{Bozek:2013sda}
P.~Bozek, A.~Bzdak, V.~Skokov, Phys.Lett., {\bf B728} (2014) 662.

\bibitem{Bozek:2013df}
P.~Bo\.zek, W.~Broniowski, Phys. Lett., {\bf B720} (2013) 250.

\bibitem{Adare:2013piz}
A.~Adare, {\it et~al.}, PHENIX Collaboration, Phys. Rev. Lett., {\bf 111}
  (2013) 212301.

\bibitem{Chatrchyan:2013eya}
S.~Chatrchyan, {\it et~al.}, CMS Collaboration, 1307.3442.

\bibitem{Basar:2013hea}
G.~Basar, D.~Teaney, 1312.6770.

\bibitem{Sickles:2013mua}
A.~M. Sickles, PHENIX Collaboration, 1310.4388.

\bibitem{Broniowski_pub:2013dia}
W.~Broniowski, E.~R. Arriola, Phys.Rev.Lett., {\bf 112} (2014) 112501.

\bibitem{blatt19521952theoretical}
J.~Blatt, V.~Weisskopf, {\em Theoretical nuclear physics}, New York: John Wiley
  and Sons, 1952.

\bibitem{brink2008history}
D.~Brink, {\em History of cluster structure in nuclei}, in: Journal of Physics:
  Conference Series, Vol. 111, IOP Publishing, 2008, p. 012001.

\bibitem{brink1965alpha}
D.~Brink, Proc. Int. School Enrico Fermi, Course, {\bf 36}.

\bibitem{freer2007clustered}
M.~Freer, Reports on Progress in Physics, {\bf 70}~(12) (2007) 2149.

\bibitem{ikeda2010clusters}
K.~Ikeda, T.~Myo, K.~Kato, H.~Toki, {\em Clusters in Nuclei-Vol.1}, Lecture
  Notes in Physics {\bf 818}, Springer, 2010.

\bibitem{beck2012clusters}
C.~Beck, {\em Clusters in Nuclei-Vol.2}, Lecture Notes in Physics {\bf 848},
  Springer, 2012.

\bibitem{Florkowski:2013sk}
W.~Florkowski, M.~Martinez, R.~Ryblewski, M.~Strickland, 1301.7539.

\bibitem{Heller:2011ju}
M.~P. Heller, R.~A. Janik, P.~Witaszczyk, Phys. Rev. Lett., {\bf 108} (2012)
  201602.

\end{thebibliography}

\end{document}